\documentclass[prb,twocolumn,showpacs,amsmath,amssymb]{revtex4}
\usepackage{graphicx}% Include figure files
\usepackage{dcolumn}% Align table columns on decimal point
\usepackage{bm}% bold math
\begin{document}   

\title{Symmetry classification of bond order parameters in cuprates}
\author{Roland Zeyher}
\affiliation{Max-Planck-Institut f\"ur Festk\"orperforschung,
             Heisenbergstrasse 1, D-70569 Stuttgart, Germany}

\date{\today}

\begin{abstract}
We study bond-order parameters for generalized $t$-$J$ models  
on a square lattice. Using the plane-wave limit the considered order 
parameters form basis functions for irreducible
representations of the symmetry transformations of the point group and of time 
reversal. We show that for 
instability wave vectors along the diagonals all possible basis functions
are either fine-tuned (i.e., obey restrictions beyond the requirements
of symmetry) or break time reversal symmetry and thus describe flux states. 
For instability wave vectors along the crystalline axes, corresponding to 
the observed case in underdoped cuprates, there are only three 
representations with $A_1$, $B_1$, and $E$ symmetry which do not break time 
reversal symmetry in the general case. We suggest that one of them has 
recently been observed in resonant elastic X-ray scattering.

\end{abstract}


\maketitle

\section{\label{sec:intro} Introduction}
There is growing evidence from  nuclear magnetic resonance \cite{Wu},
resonant X-ray scattering and diffraction 
\cite{Ghiri,Chang,Achkar,Blackburn,Blanco,Le Tacon,Comin2} and
scanning tunneling microscopy \cite{Comin1,Silva,Comin2} that 
charge ordered states play an important role
in underdoped cuprates. In particular, a charge-modulated state
with 4 incommensurate wave vectors along the crystalline axes was
detected in YBCO and in Bi-based single and double layer compounds.
Resonant elastic X-ray scattering showed that the charge order 
emerges just below the opening of the pseudogap \cite{Timusk} 
in underdoped Bi2201 \cite{Comin1}. The formation of the pseudogap, 
Fermi pockets, the appearance of quantum oscillations \cite{Doi} 
and of charge order 
may thus be intimately related in these systems. It is the aim of this 
paper to characterize charge ordered states in interacting 
systems independently of the strength of the interaction. Furthermore, the 
implications of point group and time reversal symmetries as well as the 
Hermiticity of the Hamiltonian for the order parameter (OP) will
be taken into account in greater detail than in previous treatments. 

The microscopic form of the charge OP in cuprates 
is not clear at present. To illustrate this let us consider
the following model Hamiltonian for electrons on a
square lattice which generally is believed to be relevant for 
cuprates \cite{Lee},  
\begin{eqnarray}
H = -\sum_{i,j} t_{ij} c^\dagger_{i,\alpha}c_{j,\alpha} +\;
\frac{J}{4}\sum_{{\langle i,j \rangle}} c^\dagger_{i,\alpha}
c_{i,\beta} \;\;c^\dagger_{j,\beta} c_{j,\alpha} \nonumber \\
+\frac{V}{2} \sum_{{\langle i,j \rangle}} \; 
c^\dagger _{i,\alpha} c_{i,\alpha} c^\dagger_{j,\beta}c_{j,\beta}.\;\;\;\;\;\;
\label{H}
\end{eqnarray}
$t_{ij}$ denotes the hopping amplitude of electrons between the lattice
site $i$ and $j$.  $c^\dagger_{i,\alpha}$ and $c_{i,\alpha}$ 
are fermionic creation and annihilation operators, $\alpha$ spin
indices and repeated spin indices are always summed over.
The second and third terms in Eq. (\ref{H}) describe
antiferromagnetic and Coulomb interactions between electrons on
neighboring sites $i$ and $j$ with coupling constants $J$ and
$V$, respectively. 
If double occupancies of sites are excluded Eq. (\ref{H}) represents the
well-known $t$-$J$ model for $V = -J/2$. 

The interaction terms in Eq. (\ref{H}) give rise to two kinds of charge 
OPs. A Hartree-like contraction of the third term yields an
OP proportional to   
$\langle c^\dagger_{i\alpha} c_{i\alpha} \rangle$ 
describing the charge on the site $i$. It may vary from site to site
and represents a conventional charge density wave (CDW) state. 
The exchange contractions 
of the second and third term in Eq. (\ref{H}) yield an OP
proportional to $\langle c^\dagger_{i\alpha} c_{j\alpha} \rangle$, where
$i$ and $j$ are nearest neighbor sites. This state may be called a nonlocal
CDW or a bond-order wave (BOW) state \cite{Horsch} where the CDW acquires an 
internal degree
of freedom because the electron and hole occupy different sites.
It has been shown that in the large N limit 
of the $t$-$J$ model (which corresponds to 
enforcing the constraint of no double occupancies of sites only globally) 
the phase diagram consists in the underdoped regime
of incommensurate BOW states (at zero doping of the staggered 
flux phase \cite{Affleck} as a special case) \cite{Morse,Cappelluti}. 
At the same time the conventional CDW OP is zero showing that both kinds of
charge order are independent from each other. More recently
the BOW state has been studied theoretically in more 
detail \cite{Chakravarty,Bejas,Sachdev,Allais,Wang}. 
Also models with more than one 
band \cite{Kampf} or more complex OPs \cite{Wang,Efetov} have been considered.
Recently a microscopic form for the OP
in underdoped YBCO and Bi2201 was proposed \cite{Comin2} based on
experimental data from resonant X-ray scattering.

Throughout the paper we will assume that the temperature is below the
transition temperature to the BOW state. The OPs are then in general
nonzero and their symmetry properties can be studied. We will
classify possible OPs
for BOW states by exploiting point-group and 
time reversal symmetries as well as the Hermiticity of $H$. 
In the appendix it is shown that possible OPs for the ground state are
basis functions for representations of $C_{4v}$. If the ground state
is non-degenerate in the sense that it does not contain two linearly
independent OPs the representation is irreducible. If the ground state
is degenerate and satifies a two dimensional representation
this representation may be irreducible or reducible.
In the latter case it is composed of two OPs
with different symmetries and the degeneracy is not a consequence of symmetry
but of coupling constants. In the following we will confine our discussion
to OPs which form irreducible representations of $C_{4v}$ and exlude
accidential degeneracies, additional instabilities or induced higher
harmonics \cite{Tinkham}. Explicit expressions for the OPs will be given
for BOW states with four wave vectors of the form 
$(\pm q,\pm q)$ and the form $(\pm q ,0)$ and $(0,\pm q)$. 

\section{Definition and transformation properties of the order parameter}

From Eq. (\ref{H}) follows that the BOW OP
has the form of a coupling constant times the matrix
element $\langle c^\dagger_{i,\alpha} c_{j,\alpha} \rangle$, where
$i$ and $j$ are nearest neighbors.
To simplify the nomenclature we will call the modulated part of 
$\langle c^\dagger_{i,\alpha} c_{j,\alpha} \rangle$
OP in the following. After a Fourier transform we obtain, 
\begin{equation}
\langle c^\dagger_{i,\alpha} c_{j,\alpha} \rangle = 
\sum_{{\bf k},{\bf q}} \langle c^\dagger_{{\bf k}+{\bf q},\alpha} c_{{\bf k},\alpha}
\rangle e^{-i{\bf q}{\bf r}_i} e^{-i({\bf r}_i-{\bf r}_j)\cdot {\bf k}}.
\label{Fourier}
\end{equation}
${\bf r}_i$ is the vector from the origin to the lattice site $i$. 
The sum over ${\bf q}$ in Eq. (\ref{Fourier}) includes in the 
plane-wave limit only the 
wave vectors corresponding to a charge instability of the normal 
state. They form a star of wave vectors $\{{{\bf q}_l}\}$.
Writing ${\bf r}_j = {\bf r}_i
+{\bf e}_j$, keeping ${\bf e}_j$ fixed and performing a Fourier 
transformation with respect to ${\bf r}_i$ we get
\begin{equation}
\sum_{{\bf r}_i} \langle c^\dagger_{i,\alpha} c_{j,\alpha} \rangle 
e^{i{\bf q}_l {\bf r}_i}
= \sum _{{\bf k}} \langle c^\dagger_{ {\bf k}+{\bf q}_l,\alpha}c_{{\bf k},\alpha}
\rangle e^{i{\bf e}_j\cdot {\bf k}} = F({\bf q}_l,{\bf e}_j).
\label{F}
\end{equation}
The functions  $F({\bf q}_l,{\bf e}_j)$ are defined by Eq. (\ref{F})
and represent our set of order parameters. 
If Umklapp terms are included the sum over ${\bf q}_l$ may 
not only include primary instability vectors of the normal state but
higher harmonics with wave vectors $\sum_l n_l {\bf q}_l$ where $n_l$ is an 
integer. They form new stars
and cause deviations from the plane-wave limit of the OP.
Because these higher harmonics are important only near the transition
to the commensurate phase we will neglect them in the following and restrict
ourselves to the plane-wave limit. 

The symmetry group of the square lattice is $C_{4v}$. Using the notation
of Ref. \cite{Tinkham} let us denote one of the 8 symmetry transformations 
by $R$.
Its action on the order parameter in Eq. (\ref{F}) can be written as
\begin{equation}
\langle c^\dagger_{{\bf R^{-1}}{\bf r}_i,\alpha} c_{{\bf R^{-1}}{\bf r}_j,\alpha} \rangle =
\sum_{{\bf k},l} \langle c^\dagger _{{\bf k}+{\bf R^{-1} q}_l,\alpha} c_{{\bf k},\alpha} 
\rangle
e^{-i{\bf q}_l \cdot {\bf r}_i}e^{i{\bf k}\cdot{\bf R^{-1}}{\bf e}_j},
\label{trans}
\end{equation}
where ${\bf R}$ is the 2x2 matrix representing 
$R$ in the two-dimensional direct space. After a Fourier transformation
with respect to ${\bf r}_i$ 
we find that $F({\bf q}_l,{\bf e}_j)$ transforms under $R$ into
$F({\bf R^{-1}}{\bf q}_l,{\bf R^{-1}}{\bf e}_j)$, where 
${\bf R^{-1}}{\bf q}_l$ and ${\bf R^{-1}}{\bf e}_j$ belong to the star of
wave vectors and to nearest neighbor bonds, respectively. 
This means that the set of functions $\{ F({\bf q}_l,{\bf e}_j) \}$
forms basis functions for a (reducible) representation of $C_{4v}$. 
Decomposing this reducible representation into irreducible parts
the basis function of one of the irreducible representations describes
the OP of the state corresponding to the global minimum
of the free energy. One important feature is that in general both
${\bf q}_l$ and ${\bf e}_j$ are transformed under $R$ at the same time 
and not independently from each other. This is a crucial point in our
approach. 

Further general properties of the functions $\{{ F({\bf q}_l,{\bf e}_j)}\}$
are related to time reversal
and the Hermiticity of the Hamiltonian. The operator $T$ for
time reversal is defined in real space by
\begin{equation}
T  \langle c^\dagger_{i,\alpha} c_{j,\alpha} \rangle
  =   \langle c^\dagger_{i,\alpha} c_{j,\alpha} \rangle^*,
\label{Fhilf1}
\end{equation}
where the star means conjugate complex.
Taking Fourier transforms on both sides we get,
\begin{equation}
\sum_{{\bf r}_i} e^{-i{\bf q}_l{\bf r}_i} T  \langle c^\dagger_{i,\alpha} c_{j,\alpha} \rangle =
\sum_{{\bf r}_i} e^{-i{\bf q}_l{\bf r}_i} \langle c^\dagger_{i,\alpha} c_{j,\alpha} \rangle^*,
\label{Philf2}
\end{equation}
or  
\begin{equation}
T \sum_{{\bf r}_i}  e^{i{\bf q}_l{\bf r}_i}  \langle c^\dagger_{i,\alpha} c_{j,\alpha} \rangle
 (\sum_{{\bf r}_i} e^{i{\bf q}_l{\bf r}_i} \langle c^\dagger_{i,\alpha} c_{j,\alpha} 
\rangle)^*,
\label{Philf3}
\end{equation}
or, 
\begin{equation}
T F({\bf q}_l,{\bf e}_j) = (F({\bf q}_l,{\bf e}_j))^* = 
F^*({\bf q}_l,{\bf e}j).
\label{Philf4}
\end{equation}
In Eqs. (\ref{Philf2}) and (\ref{Philf3}) $c_{j,\alpha}$ stands for
$c_{{\bf r}_i+{\bf e}_j,\alpha}$ and ${\bf e}_j$ is fixed in the sums over 
${\bf r}_i$.
$F^*({\bf q}_l,{\bf e}_j)$ is defined by the last equation. Some authors
interprete the right-hand side of Eq. (\ref{Philf2}) as a Fourier transform
of $\langle c^\dagger_{i,\alpha} c_{j,\alpha}\rangle $  
which may be written
as $(F^*)({\bf q}_l,{\bf e}_j)$. The connection to our definition 
Eq. (\ref{Philf4}) is $F^*({\bf q}_l,{\bf e}_j) = (F^*)(-{\bf q}_l,{\bf e}_j)$.
In order to avoid confusion we will always
use our definition in Eq. (\ref{Philf4}). 
Using the Hermiticity of the Hamiltonian the second half of Eq. (\ref{F})
yields 
\begin{equation}
 F^* ({\bf q}_l,{\bf e}_j) =  e^{i{\bf e}_j{\bf q}_l} F(-{\bf q}_l,-{\bf e}_j)
=  e^{i{\bf e}_j{\bf q}_l} C_2 F({\bf q}_l,{\bf e}_j).
\label{Herm}
\end{equation}
$C_2$ denotes the rotation by $\pi$. 
 
Eq. (\ref{Philf4}) implies $T^2 = 1$, which 
corresponds to the case of integral spin. \cite{Tinkham} 
Applying the Frobenius-Schur test \cite{Tinkham} to the point group
$C_{4v}$ shows that including $T$ in the set of
symmetry transformations cannot produce 
additional degeneracies of irreducible representations. Thus it is
convenient to construct first basis functions for irreducible representations
of the point group and then to check their behavior under time reversal.

\section{ Wave vectors of the BOW along the diagonals}

In the following we will first consider the case with 4 wave vectors along the
diagonals, i.e., 
${\bf q}_1=(q,q),{\bf q}_2=(-q,q),{\bf q}_3=(-q,-q),{\bf q}_4=(q,-q)$, where
$q$ lies between 0 and $\pi$. The four bond directions are denoted
by ${\bf e}_1=(1,0),{\bf e}_2=(0,1),{\bf e}_3=(-1,0),{\bf e}_4=(0,-1)$.
It is easy to see that the following 8 functions $F_l = F({\bf q}_l,{\bf e}_l),
l=1,..,4$, $F_5 = F({\bf q}_4,{\bf e}_1)$, $F_6 = F({\bf q}_2,{\bf e}_3)$,
$F_7 = F({\bf q}_1,{\bf e}_2)$, and $F_8 = F({\bf q}_3,{\bf e}_4)$ yield
a reducible representation of $C_{4v}$. Let us denote the
linear combinations of the $F_l$ which form basis functions for the
corresponding irreducible representations $\gamma$ by
\begin{equation}
F(\gamma) = \sum_{l=1}^8 c_{\gamma l}F_l.
\label{c}
\end{equation}
For the one-dimensional representations $\gamma = A_1,A_2,B_1,B_2$
one can easily determine the coefficients $c_{\gamma l}$ from the character
table of $C_{4v}$. One finds that each of these representations occurs
exactly one time, the corresponding $c_{\gamma l}$ are given in the
first 4 lines of Table 1. Using again the character table one finds
that the remaining 4 functions form 2 two-dimensional representations
$E^{(1)}$ and $E^{(2)}$. The corresponding $c_{\gamma l}$ are given in the
lines 5-8 in Table 1.     
\begin{table}
\begin{ruledtabular}
\caption{Coefficients $c_{\gamma l}$}
\vspace{0.3cm}
\begin{tabular}{c|c|c|c|c|c|c|c|c}
$\gamma$ & $F_1$ & $F_2$ & $F_3$ & $F_4$ & $F_5$ & $F_6$ & $F_7$ & $F_8$ \\
\hline
$A_1$       & 1 &  1 &  1 &  1 &  1 &  1 &  1 &  1\\
$A_2$       & 1 &  1 &  1 &  1 & -1 & -1 & -1 & -1\\    
$B_1$       & 1 & -1 &  1 & -1 &  1 &  1 & -1 & -1\\
$B_2$       & 1 & -1 &  1 & -1 & -1 & -1 &  1 &  1\\
$E^{(1)}(1)$ & 1 & -1 & -1 &  1 &  1 & -1 & -1 &  1\\ 
$E^{(1)}(2)$ & 1 &  1 & -1 & -1 &  1 & -1 &  1 & -1\\ 
$E^{(2)}(1)$ & 1 & -1 & -1 &  1 & -1 &  1 &  1 & -1\\ 
$E^{(2)}(2)$ & 1 &  1 & -1 & -1 & -1 &  1 & -1 &  1\\
\end{tabular}
\end{ruledtabular}
\end{table} 

Going back to Eq. (\ref{Herm}) we note that for the functions $F_l, l=1,..,8$
the phase factor ${\bf e}_j \cdot {\bf q}_l$ is always equal to
$q$. Multiplying Eq. (\ref{Herm}) by $c_{\gamma l}$
and summing over $l$ we obtain
\begin{equation}
F^*(\gamma) = e^{iq} C_2 F(\gamma).
\label{FF}
\end{equation}
From the character  table of $D_4$ follows that $C_2 F(\gamma)$ is equal
to $F(\gamma)$ for $\gamma = A_1,A_2,B_1,B_2$ and equal to $-F(\gamma)$
for $\gamma = E$. The solution of Eq. (\ref{FF}) is
\begin{equation}
F(\gamma) = \Big( 1+i\frac{\cos q \mp 1}{\sin q}\Bigr)Re F(\gamma).
\label{F1}
\end{equation}
where the upper sign refers to $\gamma=A_1,A_2,B_1,B_2$ and the lower sign
to $\gamma = E$, respectively. The real part of $F(\gamma)$,
$Re F(\gamma)$, may assume any real number.

For $q=0$ or $\pi$ 
the four functions $F_1,...,F_4$ form a basis for
a reducible representation of $C_{4v}$ which decomposes into $A_1$, $B_1$ and $E$
representations with the basis functions $F_1+F_2+F_3+F_4$, 
$F_1 -F_2 +F_3 -F_4$ and $F_1-F_3$, $F_2 - F_4$, respectively. Since 
Eq. (\ref{FF}) still holds the first two basis functions are real (imaginary)
and the third and fourth ones imaginary (real) for $q=0$ ($q=\pi$).  
Included as a special case is
the staggered flux phase with wave vector $(\pi,\pi)$. It has $B_1$ symmetry 
and a purely imaginary OP in agreement with previous 
conclusions. \cite{Affleck,Morse,Cappelluti,Chakravarty,Bejas}    

Let us denote the second set of 8 functions by ${\tilde F}_l,l=1,..,8$. 
Each ${\tilde F}_l$
is obtained from $F_l$ by exchanging ${\bf e}_1$
with ${\bf e}_3$ and ${\bf e}_2$ with ${\bf e}_4$. The linear space spanned
by the functions ${\tilde F}_l=1,...,8$ yields a reducible representation 
of $C_{4v}$.
Decomposing it into its irreducible parts gives one time the representations
$A_1,A_2,B_1,B_2$ and two times the representation $E$, exactly as for the
first 8 functions $F_l$. The analogue of Eq. (\ref{c}) reads
\begin{equation}
{\tilde F}(\gamma) = \sum_{l=1}^8 c_{\gamma l}{\tilde F}_l,
\label{cc}
\end{equation}
where the coefficients $c_{\gamma l}$ are the same as in Table 1.
The phase factor 
${\bf e}_j \cdot {\bf q}_l$ is for all 8 functions  equal to $-q$
so that Eq. (\ref{FF}) reads  
\begin{equation}
{\tilde F}^*(\gamma) = e^{-iq} C_2 {\tilde F}(\gamma),
\label{FFFF}
\end{equation}
with the solution
\begin{equation}
\tilde{F}(\gamma) = \Big( 1-i\frac{\cos q \mp 1}{\sin q}\Bigr)
Re \tilde{F}(\gamma).
\label{F11}
\end{equation}
$Re \tilde{F}(\gamma)$ is unrelated to $Re F(\gamma)$ and may assume any
real value. This expresses the fact that $F$ and $\tilde{F}$ describe
possible OPs for all values of $Re F(\gamma)$ and $Re \tilde{F} (\gamma)$ 

Complex functions for $F(\gamma)$ and ${\tilde F}(\gamma)$ 
do not necessarily imply that the corresponding states break $T$ 
symmetry. This is true in particular in our case because
a possible imaginary part to $F$ can come from the matrix element
but also from combinations of the exponential functions in Eq. (\ref{F}).
A general criterion for an unbroken $T$ symmetry follows from 
Eq. (\ref{Fhilf1}), namely,
\begin{equation}
T  \langle c^\dagger_{i,\alpha} c_{j,\alpha} \rangle
  =   \langle c^\dagger_{i,\alpha} c_{j,\alpha} \rangle.
\label{Fhilf5}
\end{equation}
Applying Fourier transformations on both sides of Eq. (\ref{Fhilf5})
similar as in Eqs. (\ref{Philf2}) and (\ref{Philf3}) yields
\begin{equation}
T F({\bf q}_l,{\bf e}_j) = F(-{\bf q}_l,{\bf e}_j).  
\label{dia1}
\end{equation}
Comparing with Eq. (\ref{Philf4}) yields the following criterion
which must be fulfilled if $T$ symmetry is unbroken,
\begin{equation}
F(-{\bf q}_l,{\bf e}_j) = F^*({\bf q}_l,{\bf e}_j).
\label{dia2}
\end{equation}
For the special case ${\bf q}_l=0$  the above criterion is fulfilled for 
the real OP of $A_1$ and $B_1$ symmetry, found above, but not for
the imaginary OP of the $E$ symmetry.  
Thus $T$ symmetry is unbroken for
the $A_1$ and $B_1$ and broken for the $E$ symmetry. For 
${\bf q}_l=(\pi,\pi)$ one finds that the purely
imaginary OPs of the $A_1,B_1$ symmetries break and the 
real OP of the $E$ symmetry preserves $T$ symmetry.  

Considering a general ${\bf q}_l$ along the diagonals, 
a ground state with symmetry $\gamma$ is in general given by a linear 
combination of all basis functions belonging to the same
representation. It has thus in our case the form 
$\alpha F(\gamma) + \beta {\tilde F}(\gamma)$ with coefficients
$\alpha$ and $\beta$ which have to be real to be compatible with
Eq. (\ref{F1}) and (\ref{F11}). 
Applying $T$ to this state we obtain,
\begin{equation}
T(\alpha F(\gamma) + \beta \tilde{F}(\gamma)) = \alpha T F(\gamma)
+ \beta T \tilde{F}(\gamma).
\label{Tall}
\end{equation}
Inserting the functions $F_l$ into Eq. (\ref{Philf4}), multiplying by
$c_{\gamma l}$, summing over $l$ and using Eq. (\ref{FF}) yields
\begin{equation}
T F(\gamma) = F^*(\gamma) = e^{iq}C_2 F(\gamma).
\label{dia3}
\end{equation}
Replacing $F_l$ by ${\tilde F}_l$ and using Eq. (\ref{FFFF}) instead of 
Eq. (\ref{FF}) gives
\begin{equation}
T {\tilde F}(\gamma) = {\tilde F}^*(\gamma) = e^{-iq}C_2 {\tilde F}(\gamma).
\label{dia4}
\end{equation}
Eqs. (\ref{dia3})  and (\ref{dia4}) always hold. If, in addition,
$T$ symmetry is preserved we obtain from Eq. (\ref{dia1}),
\begin{equation}
T F (\gamma) = C_2 {\tilde F}(\gamma),
\label{dia5}
\end{equation}
and
\begin{equation} 
T {\tilde F}(\gamma) = C_2 F(\gamma).
\label{dia6}
\end{equation}  
Inserting Eqs. (\ref{dia3}) - (\ref{dia6}) into Eq. (\ref{Tall}) yields 
\begin{equation}
{\tilde F}(\gamma) = e^{iq} F(\gamma).
\label{TTTT}
\end{equation}
Noting that Eqs. (\ref{F1}) and (\ref{F11}) can also be written as
\begin{equation}
F(\gamma) = e^{-iq/2}/\cos (q/2)\cdot Re F(\gamma)
\label{dia7}
\end{equation}
and 
\begin{equation}
{\tilde F}(\gamma) = e^{iq/2}/\cos (q/2)\cdot Re {\tilde F}(\gamma),
\label{dia8}
\end{equation}
Eq. (\ref{TTTT}) is equivalent to 
\begin{equation}
Re F(\gamma) = Re \tilde{F} (\gamma),
\label{F3}
\end{equation}
or, using Eqs. (\ref{c}) and (\ref{cc}), 
\begin{equation}
\sum_{l} c_{\gamma l} Re (F_l) = \sum_{l} c_{\gamma l} Re (\tilde{F}_l).
\label{sum}
\end{equation}
The functions $F_l$ and $\tilde{F}_l$ are different from each other
and the functions of each of the two sets transform within each set 
under the elements of the point group and under time reversal. Thus   
symmetry does not enforce any relation between $Re F(\gamma)$ and
$Re \tilde{F} (\gamma)$ and Eq. (\ref{F3}) will not necessarily be fulfilled
for a general Hamiltonian. 
However, this does not exclude OPs exhibiting $T$ symmetry. The 
left and right-hand sides of Eqs. (\ref{F3}) may assume independently
any real value. The case where both values are equal is not excluded and 
represents an OP with $T$ symmetry. Such a fine-tuned state may, however, be 
vulnerable
to perturbations, for instance, to a change in the coupling constants, the 
temperature etc. Since no general argument seems to exist
which protects Eq. (\ref{F3}) against such perturbations it is reasonable
to conclude that in the general case the basis functions
in Eq. (\ref{c}) and Eq. (\ref{cc}) break $T$ 
symmetry for ${\bf q}\neq 0$ and $\neq (\pi,\pi)$. Whether for a 
specific Hamiltonian the ground state breaks or preserves $T$ symmetry
can only be determined by an explicit calculation of the OPs 
and the free energy. In the next section we will encounter a totally 
different case  
where the  breaking and preserving of $T$ symmetry is enforced by symmetry
independently of the values of microscopic
coupling constants or specific Hamiltonians.  

\section{ Wave vectors of the BOW along the axes}

Next we consider the case of four wave vectors along the crystalline
axes, i.e.,
${\bf q}_1 = (q,0)$, ${\bf q}_2 = (0,q)$, ${\bf q}_3 = (-q,0)$, and
${\bf q}_4 = (0,-q)$ where $q$ lies between 0 and $\pi$. Using the previous
notation the functions $F_l,l=1,...,4$ yield irreducible representations
of $A_1,B_1$ and $E$ symmetries 
with the basis functions $F_1+F_2+F_3+F_4$, $F_1-F_2+F_3-F_4$,
$F_1-F_3$, and $F_2-F_4$, respectively. These basis functions can be 
written in the form  of Eq. (\ref{c}) where the sum runs from $1$ to $4$
and the corresponding $\gamma$ in Table 1 is chosen.
The exponential factor in
Eq. (\ref{Herm}) is in each case $e^{iq}$. Thus Eq. (\ref{FF}) holds
for these three representations and their basis functions are complex
for $q \neq 0 $. All the above results also apply to the set 
${\tilde F}_l,l=1,..,4$, if the exponential factor is replaced by $e^{-iq}$. 
Thus each of the manifolds
$F$ and $\tilde F$ lead to one $A_1,B_1$ and one $E$ representation
and their basis functions are complex for $q \neq 0$. 
The arguments concerning $T$ breaking in Eqs. (\ref{Tall}) - (\ref{sum})
can be transferred to the present star of wave vectors with the result that 
all states which are not fine-tuned in the sense discussed above 
break $T$ symmetry.

The remaining 8 functions
are conveniently split into the combinations 
$F^+_l = F({\bf q}_l,{\bf e}_{l+1}) +  F({\bf q}_l,{\bf e}_{l+3})$ and
$F^-_l = F({\bf q}_l,{\bf e}_{l+1}) -  F({\bf q}_l,{\bf e}_{l+3})$, 
where ${\bf e}_5$, ${\bf e}_6$ etc. mean ${\bf e}_1$, ${\bf e}_2$ etc. 
and $l$ runs from 1 to 4.
The functions $F^{+}_l, l=1,...,4$
lead to $A_1$, $B_1$ and $E$,
the functions $F^{-}_l,l=1,...,4$ lead to $A_2$, $B_2$ and $E$ representations. 
The corresponding basis functions are the same as for the above
$F_l$ manifold, once $F_l$ is replaced by $F^+_l$ or $F^-_l$,
respectively.
For all 8 functions the wave 
and the bond vectors are perpendicular
to each other implying that $q$ is identical to zero and that no complex 
exponential appears in Eq. (\ref{Herm}). As a result we get ${F^+}^*(\gamma)
= C_2 F^+(\gamma)$ so that $F^+(\gamma)$ is real for $\gamma = A_1,B_1$
and imaginary for $\gamma   = E$. 
Using these properties we have from Eq. (\ref{Philf4}) 
\begin{equation}
T F^+(\gamma) = C_2 F^+(\gamma),
\label{Fplus}
\end{equation}
for $l=1,...,4$. Eq. (\ref{Fplus}) is a direct consequence of the 
definition of the operator $T$ and holds in any case. If in addition 
$T$ symmetry applies Eq. (\ref{dia1}) also holds. Forming appropriate
basis functions Eq. (\ref{dia1}) is identical with Eq. (\ref{Fplus})
after taking into account that  
$F^+(A_1)$ and $F^+(B_1)$ are real and $F^+(E)$ imaginary.
This means that the condition for $T$ symmetry is automatically fulfilled
in this case for all three representations. 
Remarkable is that no condition of the kind of 
Eq. (\ref{TTTT}) or Eq. (\ref{F3}) appears which cannot be fulfilled in the 
general case. The $T$ symmetry arises here without fine-tuning and is  
enforced by symmetry.

Finally, let us consider the
basis functions $F^-(\gamma)$ of the irreducible representations 
$\gamma = A_2,B_2,E$. Eq. (\ref{Herm}) 
yields ${F^-}^*(\gamma) = C_2 F^-(\gamma)$ 
implying that $F^-(A_2)$ and $F^-(B_2)$
are real and $F^-(E)$ are imaginary. Forming basis functions in 
Eqs. (\ref{Philf4}) and (\ref{Herm}) gives
\begin{equation}
T {F^-}(\gamma) = C_2 {F^-}(\gamma).
\label{Fminus1}
\end{equation}
If in addition $T$ symmetry holds Eq. (\ref{dia1}) is fulfilled and yields
after forming linear combinations of $A_2$, $B_2$, and $E$ symmetry, 
 \begin{equation}
T F^-(\gamma) = -C_2 F^-(\gamma).
\label{Fminus2}
\end{equation}
Clearly, Eqs. (\ref{Fminus1}) and (\ref{Fminus2}) contradict each other.
Thus $T$ symmetry is always
broken in all three cases in a robust way, i.e., independent of specific
Hamiltonians and values for microscopic coupling parameters.

For completeness let us consider the case of an usual CDW without
internal bond degrees of freedom. Eq. (\ref{Herm}) reads for ${\bf e}_j=0$
$F^*({\bf q}_l) = C_2 F({\bf q}_l) = F(-{\bf q}_l)$. Thus Eq. (\ref{dia2})
is always fulfilled. Forming suitable linear combinations
to get basis functions for irreducbile representations we see that 
$T$ symmetry is always unbroken in an usual CDW.

The above analysis showed that many of the symmetry allowed OPs
break $T$ symmetry. The finite imaginary part of these OPs generate
circulating currents and space-dependent magnetic fields \cite{Affleck,Greco}
which so far could not be observed \cite{Sonier,Straessle,Mounce}.  
Concentrating therefore on $T$ conserving OPs there are none without 
fine-tuning in the case of
a star with wave vectors along the diagonals. For the experimentally
observed star with wave vectors along the crystalline axes  
there are three OPs which preserve
$T$ symmetry in a robust way. They have the symmetries $A_1$, $B_1$ and $E$ 
and originate
from the $F^+$ manifold. In particular, the $F^+(B_1)$ state with $B_1$ 
symmetry seems to be a good
candidate for the OP in underdoped cuprates \cite{Comin2}. Its 
bond charge pattern $\langle c^\dagger_{i,\alpha} c_{j,\alpha} \rangle$ is 
proportional
to $-\cos (qr_{iy})$ for ${\bf e}_j = {\bf e}_1$ and $\cos (qr_{ix})$ for
${\bf e}_j = {\bf e}_2$ and is illustrated in Fig. 1. The color on each
bond indicates the value for the corresponding bond charge. The 
pattern represents a
simple bidirectional BOW state where the charges on the hozontal and 
vertical bonds vary only in one direction. 
Different ground states for underdoped  cuprates have also been proposed,
for instance, uniaxial BOW states with \cite{Wang} or without \cite{Comin2,Wang}
$T$ breaking. Interesting is that the relevant OPs of the 
hot spot model of Ref. \cite{Wang} are closely related to our manifolds
$F^+$ and $F^-$ concerning point group and time reversal symmetries. 
%%%%%%%%%%%%%%%%%%%%%%%%%%%%%%%% FIG. 1%%%%%%%%%%%%%%%%%%%%%%%%%%%%%%%%%%%%
\begin{figure} 
\vspace*{-0ex}
\includegraphics[angle=0,width=7.0cm]{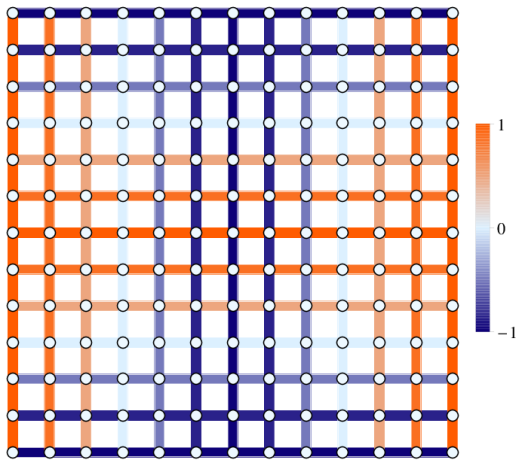}
\vspace{0.2cm}
\caption{\label{fig:1}
(Color online)
Bond charge pattern of the $F^+(B_1)$ state using $q=\pi/6$.
}
\end{figure}
%%%%%%%%%%%%%%%%%%%%%%%%%%%%%%%%%%%%%%%%%%%%%%%%%%%%%%%%%%%%%%%%%%%%%%%%%%%%

The OP deduced from experimental data in Ref. \cite{Comin2} 
uses a form for the OP which is based on the approximation
\begin{equation}
F({\bf q}_l,{\bf e}_j) =  e^{-i{\bf e}_j{{\bf q}_l/2}}
\sum_{\bf k} \Delta({\bf k}) e^{i{\bf e}_j\cdot {\bf k}}.
\label{Keim}
\end{equation}
This approximation is obtained from Eq. (\ref{F}) by
shifting the sum over ${\bf k}$ by $-{\bf q}_l/2$ and using for the matrix 
element the ${\bf q}_l$ independent 
function $\Delta({\bf k})$. From Eq. (\ref{Herm}) follows that 
$\Delta({\bf k})$ has to be real. Inserting
Eq. (\ref{F}) into Eq. (\ref{Fourier}) and using Eq. (\ref{Keim})
yields
\begin{equation}
\langle c^\dagger_{i,\alpha} c_{j,\alpha} \rangle = 
\sum_l e^{-i{\bf q}_l({\bf r}_i+{\bf r}_j)/2} \sum_{\bf k} e^{i{\bf e}_j{\bf k}}
\Delta({\bf k}).
\label{pattern}
\end{equation}
The sum over ${\bf k}$ is only nonzero if $\Delta({\bf k})$ is a linear
combination of $\cos k_x + \cos k_y$, $\cos k_x - \cos k_y$, 
$\sin k_x$ and $\sin k_y$.
Inserting these functions into Eq. (\ref{pattern}) and using the 
transformation rule of Eq. (\ref{trans}) shows that these
patterns have $A_1$, $B_1$, and $E$ symmetries. For instance, for the 
$B_1$ symmetry we have 
\begin{equation}
\langle c^\dagger_{i,\alpha} c_{j,\alpha} \rangle = 
\sum_l e^{-i{\bf q}_l({\bf r}_i+{\bf r}_j)/2} \sum_{\bf k} e^{i{\bf e}_j{\bf k}}
\Delta_0 (\cos k_x - \cos k_y),
\label{patternB1}
\end{equation}
where  $\Delta_0$ is a real constant. 
The bond charge patterns of Eq. (\ref{pattern}) are invariant if
${\bf q}_l$ is transformed as
${\bf R}^{-1} {\bf q}_l$ and ${\bf e}_j$ is kept fixed. 
This transformation describes a permutation of the wave vectors of the BOW and
does not correspond to an element of the point group $C_{4v}$.
Invariance under this transformation represents an
additional symmetry which we will call $Q$ symmetry in the following.
Applying $T$ to Eq. (\ref{pattern}) yields 
\begin{equation}
T \langle c^\dagger_{i,\alpha} c_{j,\alpha} \rangle = \pm
\langle c^\dagger_{i,\alpha} c_{j,\alpha} \rangle,
\label{TKeim}
\end{equation}
where the upper and lower sign holds for the $A_1,B_1$ and $E$ representations,
respectively. 
The corresponding basis functions are therefore real or imaginary. 
Eq. (\ref{pattern}) is identical with Eq. (S9) in 
Ref. \cite{Comin2}. This equation was used to analyse inelastic 
X-ray data in underdoped cuprates and it was concluded that, disregarding
uni-directional modulations, the ground state has $B_1$ symmetry. 
\cite{Comin2}.     
The corresponding bond charges are    
proportional to $-\cos (q(r_{ix}+1/2))-\cos (qr_{iy})$ for 
${\bf e}_j= {\bf e}_1$ and to $\cos (qr_{ix})
+\cos (q(r_{iy}+1/2))$ for ${\bf e}_j= {\bf e}_2$,
and yield a quite different pattern from that shown in Fig. 1. 

As shown above the approximation Eq. (\ref{Keim}) 
yields for each of the symmetries $A_1$, $B_1$, and $E$
just one OP. The charge patterns with $A_2$ and $B_2$ symmetries,
discussed in section IV, no longer exist in this approximation.
Eq. (\ref{Keim}) also implies   
severe restrictions in the space of OPs. For instance, if ${\bf q}_l$
and ${\bf e}_j$ are perpendicular to each other the functions
$F({\bf q}_l,{\bf e}_j)$ become identical for $l \neq j$. Moreover,
using the approximation Eq. (\ref{Keim}),  Eqs. (\ref{TTTT}) and
(\ref{F3}) hold which means $T$ symmetry for the $A_1$
and $B_1$ states and at the same time fine-tuning of OPs. 
It seems therefore preferable not to
specialize $F({\bf q}_l,{\bf e}_j)$ as in Eq. (\ref{Keim}) but to 
stick to the general form of the OPs and to use our previous general symmetry 
classification. We will restrict
the discussion in the following to $B_1$ states, but similar arguments
also apply to the symmetries $A_1$ and $E$. 

In section IV we found that there are 3 different representations with 
$B_1$ symmetry.
The ground state OP $F_0(B_1)$ will therefore be in general a linear 
combination of their basis functions, i.e., 
\begin{equation}
F_0(B_1) = \alpha F(B_1) + \beta {\tilde F}(B_1) + \delta F^+(B_1),
\label{F0}
\end{equation}
where $\alpha,\beta$ and $\delta$ are real numbers. The charge
patterns form a two-fold manifold which is quite different from
the case where Eq. (\ref{Keim}) holds and only one OP exists. This
difference is due to the constraints in the space of OPs introduced
by the approximation  Eq. (\ref{Keim}). Next we simplify $F_0(B_1)$ by
requesting that it exhibits $Q$ symmetry. $F_0(B_1)$ then specializes 
unambiguously to $\hat{F}(B_1)$ given by 
\begin{equation}
{\hat F}(B_1) = \sum_{{\bf k},l,j} \langle c^\dagger_{{\bf k}+{\bf q}_l,\alpha}
c_{{\bf k},\alpha} \rangle e^{i{\bf k}{\bf e}_j}(-1)^{j+1}.
\label{Fdach}
\end{equation} 
Regrouping the terms to form irreducible basis functions we get 
\begin{equation}
{\hat F}(B_1) = F(B_1) + {\tilde F}(B_1) - F^+(B_1). 
\label{F10}
\end{equation}
Although ${\hat F}(B_1)$ and  Eq. (\ref{patternB1})
possess both $Q$ and $B_1$ symmetry and are unambiguously determined
by these symmetries they are different. This can be seen from their
behavior under time reversal. Eq. (\ref{patternB1}) is $T$ symmetric
according to Eq. (\ref{TKeim}). Applying $T$ to Eq. (\ref{F10}) 
and using Eqs. (\ref{Tall}) - (\ref{dia5}) gives
\begin{equation}
T \hat{F}(B_1) =  e^{-iq} {\tilde F}(B1) +  e^{iq} F(B1) - F^+(B_1).
\label{That}
\end{equation}
Thus $\hat{F}(B_1)$ breaks in general $T$ symmetry because     
Eq. (\ref{TTTT}) or the equivalent Eq. (\ref{sum}) are 
in general not fulfilled so that the first two terms on the right-hand side
of Eq. (\ref{That}) are not equal to $\tilde{F}(B_1)+F(B_1)$.
According to our previous discussion
equations like $Re F(B_1) = Re {\tilde F}(B_1)$ are satisfied only for
fine-tuned OPs which neglect the contribution from circulating currents 
associated with $T$ breaking of $F(B_1), \tilde{F}(B_1)$, and $\hat{F}(B_1)$. 
In contrast to that $F^+(B_1)$ is
$T$ symmetric without any restrictions. Besides of the most 
general OP of Eq. (\ref{F0}) there are 
three distinguished and simple possibilities for the ground state OP 
with $B_1$ symmetry:\\
(a) One is $F^+(B_1)$ obeying $T$ but not $Q$ symmetry;\\
(b) Another is given by Eqs. (\ref{Fdach}) and (\ref{F10}) exhibiting $Q$
but breaking in general $T$ symmetry;\\
(c) State (b) with $T$ symmetry due to fine-tuning; this state is equivalent 
to Eq. (\ref{patternB1}).\\
Note that $Q$ symmetry is in our case not an exact symmetry because the 
point group transformations change both the momenta of the BOW and the bonds
at the same time. Thus (b) and (c) represent approximate states.   
In contrast to that state (a) is not $Q$ symmetric but necessarily invariant 
under time reversal
because this is a result of symmetry. Moreover, it is the only state which
has this property. Since experiments seem to rule out circulating currents 
in underdoped
cuprates \cite{Sonier,Straessle,Mounce} the ground state should be state (a)
if fine-tuned states (i.e., states with restrictions  
not enforced by symmetry) can be ruled out.  

\section{conclusions}
In conclusion, we have identified symmetry allowed bond OPs
for any model with nearest neighbor interactions such as the $t$-$J$ 
model and studied their properties, in
particular, with respect to time reversal. The obtained results
are relevant for recently observed charge-ordered states in underdoped
cuprates and their symmetries. The proposed OPs are more general than the
variational Ans\"atze used in the past both in theoretical and 
experimental studies. Being based on rigorous group theoretical 
considerations our results are useful to design improved
variational forms for the OP in microscopic calculations or to  
interprete experimental data.\\

\noindent{\bf Acknowledgements}
The author is grateful to H. Yamase, P. Horsch, W. Metzner, A. Greco 
and B. Keimer for useful discussions and to T. Holder for help in
producing the figure. Interesting discussions and exchanges of e-mails
with S. Sachdev, A. Allais and J. Bauer are also acknowledged.

\appendix
\section{Reducible and irreducible representations of the ground state.}

It is well known that ground state wave functions form, disregarding 
accidential degeneracies, basis functions for irreducible 
representation of the symmetry
transformations commuting with the Hamiltonian. \cite{Tinkham} 
In this appendix we will
study the question whether a similar statement is true for the ground state
of a system described by a free energy functional and OPs.

Let us denote by $F_0$ one of the OPs describing the ground state,  
i.e., which correspond to the minimum of the free energy. $F_0$
can be represented as a linear combination of the  
$F({\bf q}_l,{\bf e}_j)$ and thus transforms in a well-defined way under
point group transformations. Considering $C_{4v}$ and applying its n=8 
transformations $P_R$
to $F_0$ we denote by $R_i, i=1,...,d$ the transformations which 
lead to linearly independent functions $f_i = P_{R_{i}}F_0,\; i=1,...,d$. 
The functions $P_{R_{i}}F_0, i=d+1,...,n$ can be written as linear
combinations of the functions $f_i$. Denoting the n point group operators
by $P_{R_{\alpha}},\alpha=1,...,n$, we can write
\begin{equation}
P_{R_{\alpha}} F_0 = \sum_{i=1}^d f_i L_{i,\alpha},
\end{equation}
where $L_{i,\alpha}$ is a matrix with d rows and n columns. 
Applying $P_{R_{\alpha}}$ to $f_i$ gives 
\begin{equation}
P_{R_{\alpha}} f_i  = P_{R_{\alpha}} P_{R_{i}} F_0 = P_{R_{\alpha}R_i}F_0 = \sum_{j=1}^d f_j L_{j,\bar{i}}.
\label{P1}
\end{equation}
$\bar {i}$ denotes the column of $L$ which corresponds to the 
point group transformation $R_{\alpha} \cdot R_i$. For a fixed $R_{\alpha}$ $\bar {i}$
runs over d columns of $L$ which can be used to form a square matrix
$\Gamma(R_{\alpha})$ with $d$ rows and $d$ columns. Eq. (\ref{P1}) can now be written
as
\begin{equation}
P_{R_{\alpha}} f_i = \sum_{j=1}^d f_j \Gamma(R_{\alpha})_{j,i} .
\label{an1}
\end{equation}
Considering a product of two transformations $R_{\alpha}$ and $R_{\beta}$ we have
\begin{eqnarray}
P_{R_{\alpha}R_{\beta}} f_i = P_{R_\alpha} P_{R_\beta} f_i = 
P_{R_{\alpha}} \sum_{j=1}^d f_j \Gamma(R_{\beta})_{j,i} = \nonumber \\
\sum_{j,k} f_k \Gamma(R_{\alpha})_{k,j} \Gamma(R_{\beta})_{j.i} = \sum_{k} f_k 
(\Gamma(R_{\alpha})\cdot \Gamma(R_{\beta}))_{k,i}.
\label{produkt}
\end{eqnarray}
On the other hand is
\begin{equation}
P_{R_{\alpha}R_{\beta}} f_i = \sum_j f_j \Gamma(R_{\alpha}R_{\beta})_{j,i},
\end{equation}
so that 
\begin{equation}
\Gamma(R_{\alpha}R_{\beta}) = \Gamma(R_{\alpha}) \cdot \Gamma(R_{\beta}).
\label{an2}
\end{equation}
Eqs. (\ref{an1}) and (\ref{an2}) establish that the
matrices $\Gamma(R_{\alpha})$ together with the basis functions 
$f_i$ form a representation of $C_{4v}$. 

Assuming that no accidential degeneracy of the ground state is present 
the set of functions ${f_i}$ form a basis for the degenerate OPs of the 
ground state. If
$d=1$ there is only one linearly independent OP describing the ground
state. Moreover, this OP must form a basis function for a one-dimensional 
irreducible representation of $C_{4v}$. Thus if the ground state is 
non-degenerate its OP must be one of the basis functions belonging to 
irreducible one-dimensional representations discussed in sections III and IV.

In the case $d=2$ there are 
two linearly independent OPs describing the degenerate ground state. They 
form a two-dimensional representation of $C_{4v}$. The two basis
functions have the same free energy because they
transform into each other by symmetry operations of the point group.
Determining
the trace of the associated representation matrices ${\Gamma}_{i,j}$
there are two cases possible. The representation is irreducible and 
the two basis functions transform according to the $E$ representation.
Or, the representation is reducible and a superposition of two 
different one-dimensional representations. In this case the
free energy of the two-dimensional reducible representation may be lower
or higher than that of the irreducible representations. This means that
the case is not excluded that 
a two-fold degenerate ground state transforms according to
a reducible and not an irreducible representation.  

%$A_1$ and $B_1$. The two basis functions can be written in the form
%$\phi_{A_1} \pm \phi_{B1}$  
%Decomposing
%the reducible representation into irreducible ones by a similarity
%transformation would, however, not decrease the free energy  because the initia%l
%basis functions corresponded already to the lowest free energy. The energy
%of the two-dimensional reducible representation may, however, lie below or
%above the two, in general different, free energies of the one-dimensional
%representations $\phi_{A_1}$ and $\phi_{B1}$. Thus the state with the
%lowest free energy may switch from a two-dimensional, reducible
%to a one-dimensional representation with the same basis functions
%just by varying coupling constants. The ground state
%thus may be degenerate or non-degenerate and the degeneracy is not enforced 
%by symmetry but its existence depends on values of the 
%microscopic coupling constants.

In order to illustrate the above statements we consider a simple 
free energy model without bond degrees of freedom and
two real OPs,
\begin{equation}
\phi_1 = F({\bf q}_1 + F(-{\bf q}_1),
\end{equation}
\begin{equation}
\phi_2 =  F({\bf q}_2 + F(-{\bf q}_2).
\end{equation}
${\bf q}_1$ and ${\bf q}_2$ are wave vectors of equal length 
along the $x$ and $y$ axis, respectively. Forming the combinations
\begin{equation}
\phi_A = (\phi_1 + \phi_2)/\sqrt{2},\;\;
\phi_B = (\phi_1 - \phi_2)/\sqrt{2},
\end{equation}
yields basis functions $\phi_A$ and $\phi_B$ for a $A_1$ and a $B_1$ 
representation, respectively. 
We consider the following free energy functional, 
\begin{equation}
F = \frac{a}{2} (\phi_A^2 + \phi_B^2) +\frac{1}{4} \phi_A^4 
+ \frac{g_B}{4} \phi_B^4 +\frac{g}{2} \phi_A^2 \phi_B^2,
\label{FA}
\end{equation}
in the parameter range $0<g_B < 1$ and $g \geq -\sqrt{g_B}$.
The underlying symmetry group is $C_{4v}$. The coefficient $a$ is proportional 
to $T-T_c$ and becomes negative below
the transition temperature $T_c$ to the BOW state. Note that the prefactor
$a$ is the same for both order parameters though the
latter have different symmetries. In the usual terminology this corresponds to
an accidential degeneracy. In our case this degeneracy is caused by the fact 
that the diverging susceptibilities in the normal state 
at $\pm {\bf q}_1$ and $\pm {\bf q}_2$ are related by
point group operations. There is only one $T_c$ for both 
symmetries $A_1$ and $B_1$. Below $T_c$ the OPs $\phi_A$ and $\phi_B$  
will become finite and in general be also different due to
the anharmonic terms in $F$.

Solving the extremal equations for $F$ yields the following 
solutions: 
\begin{equation} 
(a)\; \phi_A=0, \phi_B = \pm \sqrt{\frac{-a}{g_B}}, \hspace*{2cm}
\label{solu1}
\end{equation}
\begin{equation}
(b)\; \phi_B=0, \phi_A = \pm \sqrt{-a}, \hspace*{2cm} 
\label{solu2}
\end{equation}
\begin{equation}
(c)\; \phi_A = \pm \sqrt{\frac{(-a)(g_B-g)}{g_B-g^2}},
   \phi_B = \pm \sqrt{\frac{(-a)(1-g)}{g_B-g^2}}.
\label{solu3}
\end{equation}
The free energy Eq. (\ref{FA}) is invariant under $\phi_A \rightarrow
-\phi_A$ and $\phi_B \rightarrow -\phi_B$ which leads to  
additional degeneracies described by $\pm$ in Eqs. (\ref{solu1}) - 
(\ref{solu3}). It is convenient to take this degeneracy tacitly into
account and to consider only half of the above solutions. 
Calculating also the corresponding free energies we find
that (a) represents  a basis function $\phi_B$ of length 
$\sqrt{\frac{-a}{g_B}}$ for a 
$B_1$ representation with energy $ -\frac{a^2}{4g_B}$.
Similarly, (b) represents a basis function of length  $\sqrt{-a}$ for a
$A_1$ representation with energy $-\frac{a^2}{4}$. Finally, (c) consists of 
two degenerate basis functions with the components 
$(\phi_A,\phi_B)$ and $(\phi_A,-\phi_B)$,
respectively, where $\phi_A$ and $\phi_B$ are given by the expressions of
Eq. (\ref{solu3}), omitting $\pm$ in these expressions. The two   
basis functions yield a reducible two-dimensional representation 
containing both $A_1$ and $B_1$ symmetries. The corresponding free energy is 
$-\frac{a^2(1+g_B-2g)}{4(g_B-g^2)}$.
%%%%%%%%%%%%%%%%%%%%%%%%%%%%%%%% FIG. 2%%%%%%%%%%%%%%%%%%%%%%%%%%%%%%%%%%%%
\begin{figure} 
\vspace*{-12ex}
\includegraphics[angle=270,width=8.0cm]{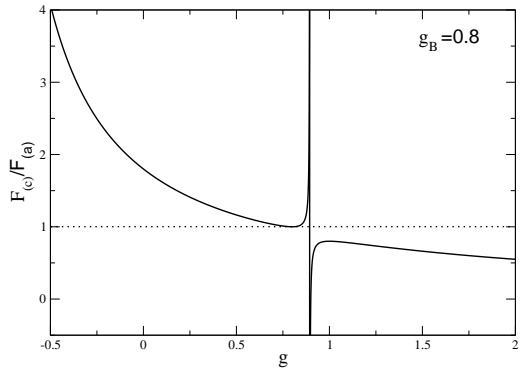}
\vspace{0.2cm}
\caption{\label{fig:2}
%(Color online)
Ratio of the free energies $F_{(c)}$ and $F_{(a)}$ as a function of $g$
using $g_B = 0.8$. The dashed line corresponds to a ratio of 1.
}
\end{figure}
%%%%%%%%%%%%%%%%%%%%%%%%%%%%%%%%%%%%%%%%%%%%%%%%%%%%%%%%%%%%%%%%%%%%%%%%%%%%

Because of the assumption $0 < g_B <1$ the free energy of (a) is always
lower than that of (b) so that the ground state is given either by (a) or 
by (c).
Fig. 2 shows the ratio of the free energies of (c) and (a) as a function 
of $g$ for $g_B = 0.8$. For $g > \sqrt{g_B}$ the curve in Fig. 2 is
always below 1. Thus the solution (a) has in this region the lowest free
energy and describes the stable state.  
It is non-degenerate and its basis function belongs to 
an irreducible representation of $B_1$ symmetry. For $|g| < \sqrt{g_B}$ the 
curve in Fig. 2 is larger or equal to one. As a result   
solution (c) has the lowest free energy in this interval 
and describes a degenerate ground 
state. It is given by a two-dimensional reducible
representation which consists both of $A_1$ and $B_1$ components. 
Our calculation shows that this two-dimensional reducible representation 
may have a lower free energy than the $A_1$ and $B_1$ components.
Our calculation also demonstrates that the same basis functions
may describe a degenerate or a non-degenerate ground state depending
on the values of the coupling constants. This means that the degeneracy of a
ground state described by a reducible two-dimensional representation is not 
enforced 
by symmetry as in the case of irreducible representations but depends in general
on the values of the coupling constants. It thus can be 
considered to be accidential. 
%Accepting this wide notion of accidential
%symmetry the first sentence of this appendix also applies to the ground
%state of systems described by a free energy functional in terms of OPs.
%Such a case cannot occur in Quantum
%Mechanics where the Hamiltonian and the vectors of the Hilbert space 
%correspond to the free energy and the OPs of our case. In this case
%a superposition of two vectors with different symmetries and energies
%will have always a larger energy expectation value than that of one of the 
%pure states. This difference is mainly caused by the fact that   
%$H$ is a linear operator in Hilbert space in contrast ot the free
%energy in the space of OPs.
  
%%%%%%%%%%%%%%%%%%%%%%%%%%%%%%%%%%%%%%%%%%%%%%%%%%%%%%%%%%%%%%%%%%%%%%%%%%%%
%%
%%                           REFERENCES
%%
%%%%%%%%%%%%%%%%%%%%%%%%%%%%%%%%%%%%%%%%%%%%%%%%%%%%%%%%%%%%%%%%%%%%%%%%%%%%

\end{document}